\documentstyle[preprint,tighten,aps]{revtex}
\begin{document}
\preprint{\vbox{ \null\hfill INFNCA-TH9520 \\
\null\hfill DFTT 58/95 \\
\null\hfill hep-ph/9510231}}
\vskip2.5truecm
\draft
\title{$\protect\bbox{\chi_{c2}\to\rho\rho}$ and the
 $\protect\bbox{\rho}$ polarization in massless
 perturbative QCD: \\ how to test the distribution amplitudes}

\author{Mauro Anselmino}

\address{Dipartimento di Fisica Teorica, Universit\`a di Torino and \\
      INFN, Sezione di Torino, Via P. Giuria 1, I--10125 Torino, Italy}

\author{ Francesco Murgia}

\address{Istituto Nazionale di Fisica Nucleare,
Sezione di Cagliari \\
via Ada Negri 18, I--09127 Cagliari, Italy}

\date{October 1995}

\maketitle

\begin{abstract}
We compute the helicity density matrix of $\rho$ vector
mesons produced in the two-body decays of polarized
$\chi_{c2}$'s in the framework of massless perturbative QCD.
The $\chi_{c2}$'s are either exclusively or inclusively
produced in $p \, \bar p$ or $p \, p$ interactions,
via quark-antiquark annihilation or gluon fusion.
Our results show unambiguous significant differences depending
on the choice of the $\rho$ distribution amplitudes and allow
to discriminate between different proposed models.

\end{abstract}

\pacs{ PACS numbers: 13.25.Gv, 12.38.Bx, 13.88.+e, 14.40.Gx}

\narrowtext

\section{Introduction}
\label{intro}

The general theoretical framework for the description
of high energy exclusive processes within perturbative QCD
has been widely discussed in the literature \cite{bro} and,
although it has not yet reached the same level of reliability
and variety of applications as the factorization
theorem for the inclusive interactions, is expected
to work at large momentum transfers and indeed several
successful calculations have been performed so far in some
simple cases \cite{bro,che}. The numerical results
depend strongly on the internal structure of the hadrons
involved which, in the exclusive cases, are described by
the hadron wave functions or distribution amplitudes,
analogous to the distribution functions of the
inclusive case. Perturbative QCD fixes the asymptotic
form of these amplitudes at $Q^2 \to \infty$ and their
general evolution, but in any realistic calculation they
have to be taken as phenomenological quantities, to be
experimentally determined via a set of physical information
and then used in other processes.
This procedure has led to the success of some distribution
amplitudes motivated by QCD sum rules \cite{che,kin,che89,stef}.

Some criticism has been advanced to the
validity of the whole scheme, on the ground that it necessarily
involves a contribution from soft non perturbative regions
\cite{isg}; such criticism has been answered by
noticing that these soft contributions are suppressed by
Sudakov form factors \cite{ster}.
While this gives a better understanding and credibility
to the theoretical model it weakens the agreement between
theoretical computations, based on the QCD sum rule distribution
amplitudes \cite{che}, and experiment;
in fact these distribution amplitudes enhance the
contribution of the soft end point regions, which are now
known to be depressed by the Sudakov form factors.
Moreover, some lattice estimates \cite{lat1,lat2} and further
theoretical considerations \cite{rad1,rad2} have suggested different
shapes for the distribution amplitudes, closer to the asymptotic
ones; the situation is then less clear and better phenomenological
studies have to be performed in order to improve our knowledge
of the properties of the hadrons and to reach the
possibility of making genuine predictions.

We consider here the $\chi_{c2} \to \rho \, \rho$
decay process of polarized charmonium states created
in $p\,p$ or $p\,\bar p$ interactions and show how
the observation of the polarization of the vector meson,
via a measurement of its diagonal helicity density matrix
elements, neatly depends on the $\rho$ distribution
amplitudes and helps in discriminating between different
kinds of these quantities.

The same process has already been studied, with different
purposes, in two previous papers \cite{noi1,noi2},
first in the exclusive channel
$p\,\bar p \to \chi_{c2} \to \rho \, \rho$ \cite{noi1}
and then in the inclusive one
$p\,p \to \chi_{c2}+X \to \rho \, \rho +X$ \cite{noi2}.
The goal there was that of assessing the significance
of higher twist mass effects, rather than that of discussing
the role of different distribution amplitudes: actually,
it turned out that, independently of the chosen meson wave
functions, only mass effects could give origin to
non zero off-diagonal helicity density matrix elements,
$\rho_{1,-1}(\rho)$ or $\rho_{1,0}(\rho)$.
Any non zero measurement of even a single one of
these non diagonal elements, signaling the violation of
the helicity conservation rule, would be a clear
signature of mass corrections.

Here we compute, within massless perturbative QCD,
the diagonal element $\rho_{1,1}(\rho)$ and study
its dependence on the distribution amplitudes,
$\varphi(x)$; to be safe from mass corrections we
exploit the results of Refs. \cite{noi1} and \cite{noi2}
and only consider kinematical regions in which such
corrections are known to be small. The variations
resulting from the choice of different distribution
amplitudes cannot then be ascribed to other reasons.
In the next Section we briefly recall the computations
of Refs. \cite{noi1} and \cite{noi2} and show analytical
expressions which are numerically evaluated in Section 3,
which contains our results and where we also give some conclusive
comments. We improve on the computations of Refs. \cite{noi1} and
\cite{noi2} by explicitly introducing the $Q^2$ evolution of the
distribution amplitudes and the strong coupling constant $\alpha_s$.

\section{Computation and measurement of the
$\protect\bbox{\rho}$ helicity density matrix}

The full processes we are considering are either
the exclusive,
\begin{equation}
p \, \bar p \to \chi_{c2} \to \rho \, \rho \,,
\label{exc}
\end{equation}
or the inclusive,
\begin{equation}
p \, p \to \chi_{c2} + X \to \rho \, \rho + X \,,
\label{inc}
\end{equation}
production of a pair of $\rho$ vector mesons
with the subsequent decay
\begin{equation}
\rho \to \pi \, \pi \,,
\label{dec}
\end{equation}
and the quantity experimentally observed is the angular
distribution of either one of the pions in the helicity
rest frame of the decaying $\rho$ \cite{bou}.
In the inclusive case (\ref{inc}) the two initial
particles need not be protons and might be other hadrons.

The pion angular distribution depends on the spin state of
the $\rho$ via the elements of its helicity density matrix
$\rho_{\lambda,\lambda^\prime}(\rho)$,
\begin{eqnarray}
W(\Theta,\Phi) &=& {3 \over 4\pi}
\bigl[ \rho_{0,0}\cos^2\Theta + (\rho_{1,1}-
\rho_{1,-1})\sin^2\Theta\cos^2\Phi \nonumber \\
&+& (\rho_{1,1}+\rho_{1,-1})\sin^2\Theta\sin^2\Phi \nonumber \\
&-& \sqrt{2}(\text{Re}\,\rho_{1,0})\sin 2\Theta\cos\Phi \bigr] \,,
\label{ang}
\end{eqnarray}
where $\Theta$ and $\Phi$ are, respectively, the polar
and azimuthal angles of the pion as it emerges from the
decay of the $\rho$, in the $\rho$ helicity rest frame.
By integrating Eq. (\ref{ang}) over $\Phi$ or $\Theta$
one has respectively the polar and azimuthal distributions
\begin{eqnarray}
W(\Theta) &=& {3\over 2}\bigl[\rho_{0,0}+
(\rho_{1,1}-\rho_{0,0})\sin^2\Theta \bigr] \,,
\label{anp} \\
W(\Phi) &=& {1\over 2\pi}\bigr[ 1-2\rho_{1,-1}
+4\rho_{1,-1}\sin^2\Phi \bigr] \,.
\label{ana}
\end{eqnarray}
Measurements of the above angular distributions
yield direct information on
$\rho_{\lambda,\lambda^\prime}(\rho)$.

On the other hand, the vector meson helicity
density matrix $\rho(\rho)$ can be computed in
perturbative QCD \cite{bro,che}:
\begin{equation}
\rho_{\lambda,\lambda^{\prime}}(\rho;\theta,\phi) =
{1\over N} \sum_{\mu,M,M^{\prime}}
A_{\lambda,\mu;M}A_{\lambda^{\prime},\mu;M^{\prime}}^* \,
\hat \rho_{M,M^{\prime}}(\chi_{c2})
\label{rr}
\end{equation}
where $N$ is the normalization factor such that Tr$[\ \rho\ ]=1$,
\begin{equation}
N(\theta,\phi) = \sum_{\lambda,\mu,M,M^{\prime}}
A_{\lambda,\mu;M}(\theta,\phi)
A_{\lambda,\mu;M^{\prime}}^*(\theta,\phi) \,
\hat \rho_{M,M^{\prime}} \,.
\label{nor}
\end{equation}
The $A_{\lambda,\mu;M}(\theta,\phi)$'s are the
helicity amplitudes for the decay of a $\chi_{c2}$
at rest, with spin third component $J_z=M$, into
two $\rho$ vector mesons at angles $\theta, \phi$
and $\pi-\theta, \pi + \phi$ and helicities
$\lambda$ and $\mu$ respectively,
$\chi_{c2}(J_z=M) \to \rho(\lambda) \, \rho(\mu)$;
$\hat \rho_{M,M^{\prime}}(\chi_{c2})$ is the spin
density matrix of the decaying charmonium state.

Explicit expressions of the helicity decay amplitudes
$A_{\lambda,\mu;M}$ can be found in Ref. \cite{noi1},
computed within perturbative QCD in the general case
in which (constituent) quark masses are taken into account;
the massless case, which we consider here as explained
in the Introduction, is simply obtained by setting
$\epsilon =0$ in the expressions of Ref. \cite{noi1}.
These decay amplitudes contain the $\rho$ distribution
amplitudes, $\varphi(x)$.

In order to compute the $\rho$ density matrix via
Eqs. (\ref{rr}) and (\ref{nor}) we still have to
specify the spin state of the $\chi_{c2}$;
to do so we treat separately the exclusive (\ref{exc})
and inclusive (\ref{inc}) production processes.
The charmonium state is produced in the exclusive
$p \, \bar p$ channel via quark-antiquark annihilations
and the helicity conservation rule of massless
perturbative QCD entails \cite{noi1}:
\begin{eqnarray}
\hat\rho_{M,M^\prime}^{ex} &=& 0
\qquad\qquad M \not= M^\prime
\nonumber\\
\hat\rho_{0,0}^{ex} &=& \hat\rho_{2,2}^{ex} =
\hat\rho_{-2,-2}^{ex} = 0
\label{cex} \\
\hat\rho_{1,1}^{ex} &=& \hat\rho_{-1,-1}^{ex} =
1/2 \,.\nonumber
\end{eqnarray}

In the inclusive channel (\ref{inc}) the $\chi_{c2}$
is produced dominantly via two gluon fusion and,
assuming a non relativistic wave function for
the charmonium state, one finds \cite{noi2}:
\begin{eqnarray}
\hat\rho_{M,M^\prime}^{in} &=& 0 \qquad\qquad M \not= M^\prime
\nonumber\\
\hat\rho_{0,0}^{in} &=& \hat\rho_{1,1}^{in} = \hat\rho_{-1,-1}^{in} = 0
\label{cin} \\
\hat\rho_{2,2}^{in} &=& \hat\rho_{-2,-2}^{in} = 1/2 \,. \nonumber
\end{eqnarray}

By exploiting Eqs. (\ref{rr})-(\ref{cin}) and the analytical
results of  Ref. \cite{noi1} one derives for the helicity density
matrix of the $\rho$ meson, produced either via the exclusive
process (\ref{exc}) or the inclusive one (\ref{inc}):
\begin{eqnarray}
\rho_{\lambda,\lambda^\prime} &=& 0
\qquad\qquad \lambda \not= \lambda^\prime \nonumber\\
\rho_{0,0} &=& 1-2\rho_{1,1}
\label{rrf} \\
\rho_{1,1} &=& \rho_{-1,-1} =
{1\over 2}\>{\quad 1\over\displaystyle
1+3{\strut |\tilde A_{0,0}|^2
\over\displaystyle |\tilde A_{1,-1}|^2}
F(\theta)} \nonumber
\end{eqnarray}
where the reduced amplitudes
$\tilde A_{\lambda,\lambda^\prime}$ are given in
Eqs. (2.10, 11) and (2.14, 15) of Ref. \cite{noi1}
and do not depend on the $\rho$ production angles
$\theta$ and $\phi$, but do depend on the distribution
amplitudes. Eq. (\ref{rrf}) has the same form both for
exclusive and inclusive production of $\rho$, but the
dependence on the production angle $\theta$ is different
in the two cases
\begin{eqnarray}
F^{ex}(\theta) &=& {\cos^2\theta \over 1 +
 \cos^2\theta} \label{fex} \\
F^{in}(\theta) &=& {\sin^4\theta \over 1 +
 6\cos^2\theta + \cos^4\theta} \,.
\label{fin}
\end{eqnarray}

Eq. (\ref{rrf}) shows that in massless perturbative
QCD only the diagonal elements of $\rho(\rho)$ survive
and their values, which depend on the $\rho$ distribution
amplitudes, will be discussed in the next Section;
Eq. (\ref{rrf}) also implies that the azimuthal angular
distribution (\ref{ana}) of the pions emitted by the
$\rho$ vector meson is flat, $W(\Phi) = 1/(2\pi)$,
whereas the polar one (\ref{anp}) depends on $\rho_{1,1}$
(or $\rho_{0,0}$) and can be rewritten as
\begin{eqnarray}
W(\Theta) &=& {3\over 2}\bigl[\rho_{1,1}+
(1-3\rho_{1,1})\cos^2\Theta \bigr]
\label{anp1} \\
&=& {3\over 4}\bigl[2\rho_{0,0}+
(1-3\rho_{0,0})\sin^2\Theta \bigr] \,.
\label{anp2}
\end{eqnarray}

\section{Numerical evaluation of $\protect\bbox{\rho_{1,1}}$
and dependence on the distribution amplitudes}

We report here for convenience from Ref. \cite{noi1}
the explicit expression of $|\tilde A_{0,0}|/
|\tilde A_{1,-1}|$ which, together with the
$F(\theta)$ functions given in Eqs. (\ref{fex}) and
(\ref{fin}), enters the computation of the helicity
density matrix $\rho_{1,1}$ according to Eq. (\ref{rrf}):
\begin{equation}
{|\tilde A_{0,0}| \over |\tilde A_{1,-1}|} =
\frac{1}{\sqrt{6}} \left(\frac{f_L}{f_T}\right)^2
{|I_{0,0}| \over |I_{1,-1}|}
\label{ii}
\end{equation}
where
\begin{eqnarray}
I_{1,-1} &=& -{1\over 32} \int_0^1 dx dy \,
\varphi_T(x,\tilde Q^2_x) \, \varphi_T(y, \tilde Q^2_y)
\nonumber \\ & \qquad\qquad \times &
{\alpha_s \bigl[ xyM_\chi^2 \bigr] \,
\alpha_s \bigl[(1-x)(1-y)M_\chi^2 \bigr]
\over xy(1-x)(1-y)(2xy-x-y)} \label{int}
\end{eqnarray}
\begin{eqnarray}
I_{0,0} &=& -{1\over 32} \int_0^1 dx dy \,
\varphi_L(x,\tilde Q^2_x) \, \varphi_L(y, \tilde Q^2_y)
\nonumber \\ & \qquad\qquad \times &
{\alpha_s \bigl[ xyM_\chi^2 \bigr] \,
\alpha_s \bigl[(1-x)(1-y)M_\chi^2 \bigr]
\over xy(1-x)(1-y)(2xy-x-y)}
\nonumber \\ & \qquad\qquad \times &
\left[ 1 + {(x-y)^2 \over 2xy-x-y} \right] \,.
\label{inl}
\end{eqnarray}

In the above equations $M_{\chi}$ is the $\chi_{c2}$ mass,
$\tilde Q_z =$ min$(z,1-z)Q$ \cite{brod80} and $\varphi_L,
\varphi_T$ are the $\rho$ distribution amplitudes which,
following Refs.~\cite{bro,che}, are in general assumed to
be different for longitudinally ($L, \lambda = 0$) or
transversely ($T, \lambda = \pm 1$) polarized vector mesons;
$f_L$ and $f_T$ are the corresponding decay constants.

To test the dependence of the numerical values of
$\rho_{1,1}$ on the distribution amplitudes we have
performed the computations of the integrals (\ref{int})
and (\ref{inl}) choosing two typical different sets of
$\varphi_L$ and $\varphi_T$, i.e. the symmetric
distribution amplitudes
\begin{equation}
\varphi_L(x) = \varphi_T(x) = 6x(1-x)
\label{asy}
\end{equation}
and the QCD sum rule ones \cite{che}
\begin{eqnarray}
\varphi_L(x,\tilde Q^2_x) &=& 6x(1-x)
\nonumber \\ &\times &
\Biggl\{1 + {1\over 5} C^{3/2}_2(2x-1) \left[
{\alpha_s(\tilde Q^2_x) \over \alpha_s(\mu^2_L)}
\right]^{2/3} \Biggr\} \label{fil} \\
\varphi_T(x,\tilde Q^2_x) &=& 6x(1-x)
\Biggl\{ \left[
{\alpha_s(\tilde Q^2_x) \over \alpha_s(\mu^2_T)}
\right]^{4/25} \nonumber \\ & - \ &
{1\over 6} C^{3/2}_2(2x-1) \left[
{\alpha_s(\tilde Q^2_x) \over \alpha_s(\mu^2_T)}
\right]^{52/75} \Biggr\}
\label{fit}
%%%%%%%%%%%%%%%%%%%%%%
%\varphi_L(x,\tilde Q^2_x) &=& f_\pi \sqrt{3\over 2} x(1-x)
%\Biggl\{1 + {1\over 5} C^{3/2}_2(2x-1) \left[
%{\alpha_s(\tilde Q^2_x) \over \alpha_s(\tilde \mu^2_L)}
%\right]^{3/2} \Biggr\} \label{fil} \\
%\varphi_T(x,\tilde Q^2_x) &=& f_\pi \sqrt{3\over 2} x(1-x)
%\Biggl\{ \left[
%{\alpha_s(\tilde Q^2_x) \over \alpha_s(\tilde \mu^2_T)}
%\right]^{4/25} - {1\over 6} C^{3/2}_2(2x-1) \left[
%{\alpha_s(\tilde Q^2_x) \over \alpha_s(\tilde \mu^2_T)}
%\right]^{52/75} \Biggr\}
%\label{fit}
%%%%%%%%%%%%%%%%%%%%%
\end{eqnarray}
where $\mu_L^2 = 0.5$ (GeV/$c)^2$,
$\mu_T^2 = 0.25$ (GeV/$c)^2$ \cite{che}, and $C(z)$
denotes Gegenbauer polynomials. In both cases we have $f_L=f_T$.

The latter set includes explicitly the QCD $Q^2$
evolution of the distribution amplitudes \cite{bro,brod79}
and the strong coupling constant has a smooth limited
small $Q^2$ behaviour \cite{corn}
\begin{equation}
\alpha_s(Q^2) = {12\pi \over 25 \,
\ln\left[(Q^2+4m_g^2)/\Lambda^2\right]}
\label{alf}
\end{equation}
with $m_g = 0.5$ GeV/$c$ and $\Lambda = 0.2$ GeV/$c$. The effective
gluon mass $m_g$, which avoids enhancements in the soft end point
regions, models in a convenient way the effects of Sudakov
form factors \cite{ster}.

Notice that Eqs. (\ref{fil}) and (\ref{fit}), although they
imply the usual asymptotic ($Q^2 \to \infty$) behaviour \cite{bro}
\begin{equation}
\varphi_T(x, Q^2) = \varphi_L(x, Q^2) \left[
{\alpha_s(Q^2) \over \alpha_s(\mu^2_T)} \right]^{4/25} \,,
\label{asb}
\end{equation}
lead, at the moderate $Q^2$ values involved in the processes
considered here, to longitudinal and transverse distribution
amplitudes which strongly differ in their $x$ dependences;
this feature strongly affects the numerical results.

The set (\ref{asy}), on the contrary, has the same $x$
dependence both for $\varphi_L$ and $\varphi_T$ and no $Q^2$
evolution; the same $x$ dependence is the dominant feature and
we have checked that by adding to $\varphi_T$ a $Q^2$
dependent factor in agreement with Eq. (\ref{asb}) the numerical
results for the helicity density matrix do not change
significantly. This is the reason why we have only considered
the two choices of $\varphi_L$ and $\varphi_T$ given above: they
are representative of whole classes of distribution amplitudes.

In Figs. 1 and 2 we show the resulting values of
$\rho_{1,1}(\rho)$ as a function of the production
angle $\theta$ for the two choices of the distribution
amplitudes (\ref{asy}) and (\ref{fil}, \ref{fit}) and,
respectively, for the exclusive and inclusive production
processes. In both cases the two sets of distribution
amplitudes lead to clearly different results.
Notice that $\rho_{1,1}(\theta)$ is symmetric around
$\theta = \pi/2$, see Eqs. (\ref{fex}) and (\ref{fin}).

The values of $\rho_{1,1}$ fix the shape of the polar
angle distribution (\ref{anp1}) of the pion resulting
from the $\rho$ decay, which is shown in Figs. 3 and 4,
respectively for $\rho$ vector particles produced
at $\theta = 50^{\circ}$ in the exclusive case and
$\theta = 90^{\circ}$ in the inclusive one.
We have chosen such angles because in these kinematical
regions the mass corrections are negligible and cannot
affect our study of the distribution amplitudes;
this can be seen from the results of Refs. \cite{noi1,noi2}
(which, however, were derived with a fixed coupling
constant and no $Q^2$ evolution of the distribution
amplitudes) and has also been explicitly verified with
the present improvement of the calculations.
Again, the two choices of distribution amplitudes,
both in the exclusive and the inclusive channels,
lead to very different pion angular distributions,
whose observation would allow to discriminate between them.

Our study shows that different distribution amplitudes
give different results for the diagonal elements of the
$\rho$ vector meson helicity density matrix; the $\rho$'s
are decay product of charmonium states $\chi_{c2}$,
either exclusively or inclusively produced. A measurement
of $\rho_{1,1}$ is difficult and might not allow a detailed
comparison between similar distribution amplitudes, but certainly
allows to discriminate symmetric distribution amplitudes,
Eq. (\ref{asy}), or other distribution amplitudes with
$\varphi_L \sim \varphi_T$,
from those similar to the ones inspired by QCD sum
rules,  Eqs. (\ref{fil}, \ref{fit}).
Even for such a qualitative information is worth attempting the
measurements we propose here; the more we know about subtle
hadronic properties and the more we gain in predictive power.

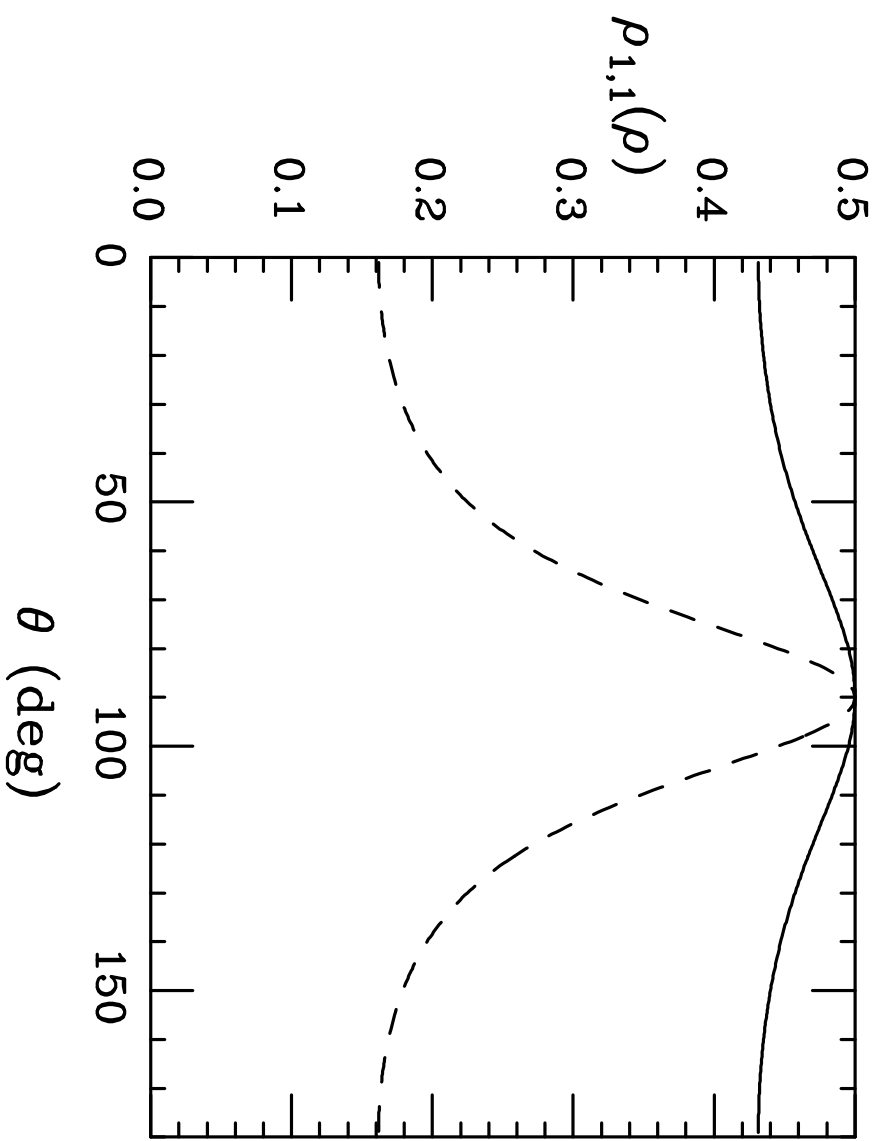
\begin{figure}
\caption[fig1]{ Values of $\rho_{1,1}(\rho)$ as a
 function of the $\rho$ meson production angle $\theta$,
 in the exclusive process (\ref{exc}). Solid curve: symmetric
 distribution amplitudes, Eq. (\ref{asy}); dashed curve: QCD sum rule
 distribution amplitudes, Eqs. (\ref{fil}, \ref{fit}). }
\label{rhex}
\end{figure}

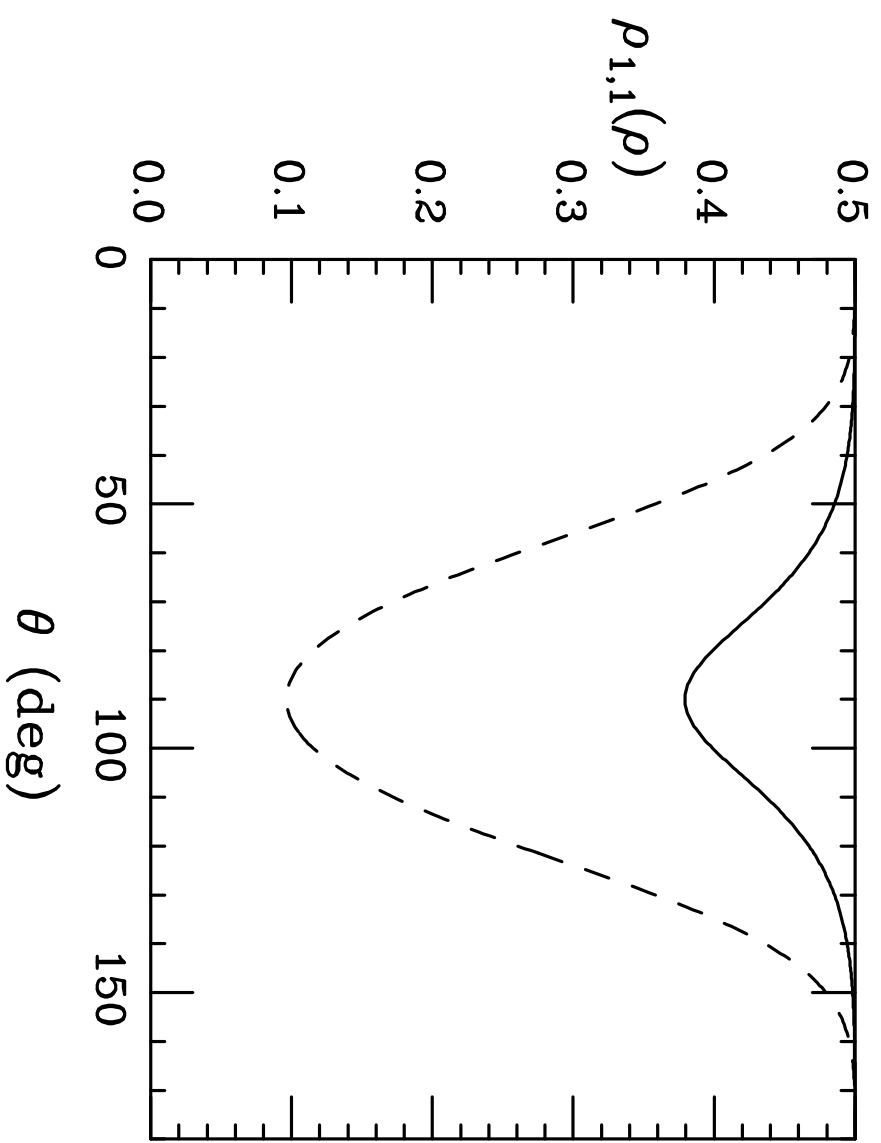
\begin{figure}
\caption[fig2]{ Values of $\rho_{1,1}(\rho)$ as a function
 of the $\rho$ meson production angle $\theta$, in the
 inclusive process (\ref{exc}). Same notations for the
 curves as in Fig. 1. }
\label{rhin}
\end{figure}

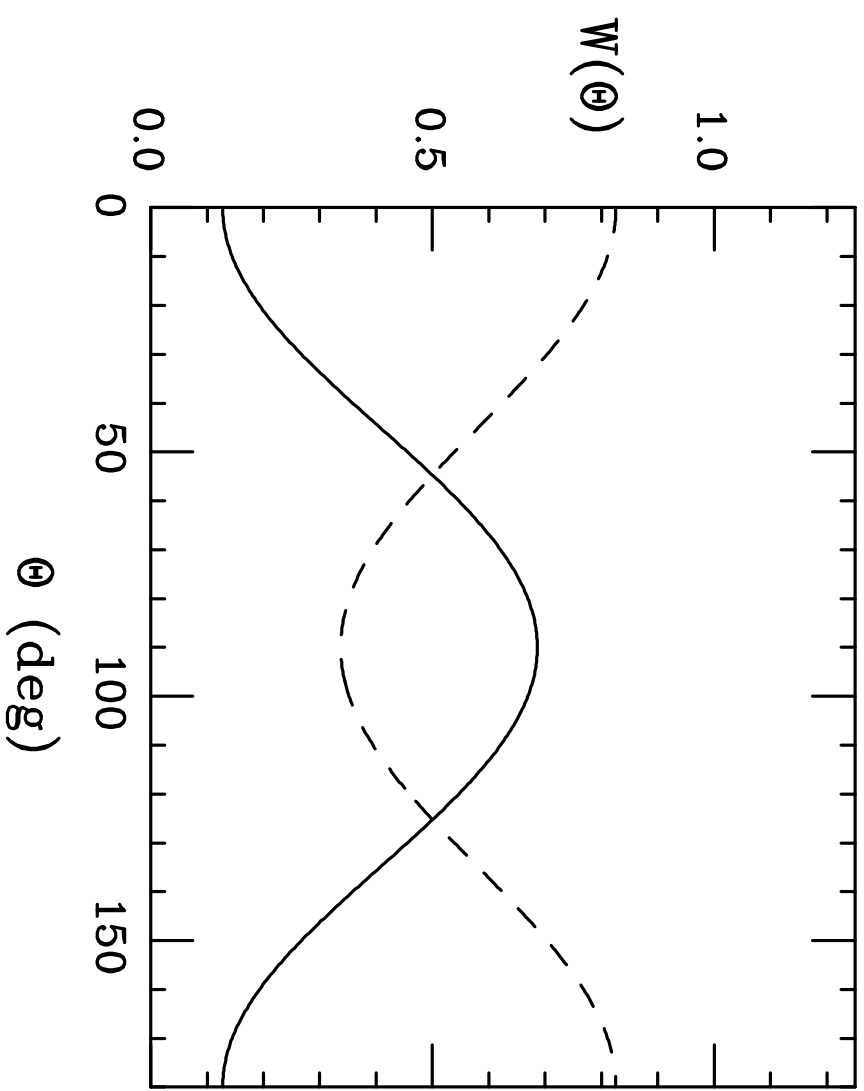
\begin{figure}
\caption[fig3]{ Plot of the angular distribution,
 $W(\Theta)$, of the $\pi$ emitted in the $\rho$ decay.
 The $\rho$ has been produced in the exclusive channel
 at an angle $\theta = 50^\circ$. Same notations for the
 curves as in Fig. 1. }
\label{wex}
\end{figure}

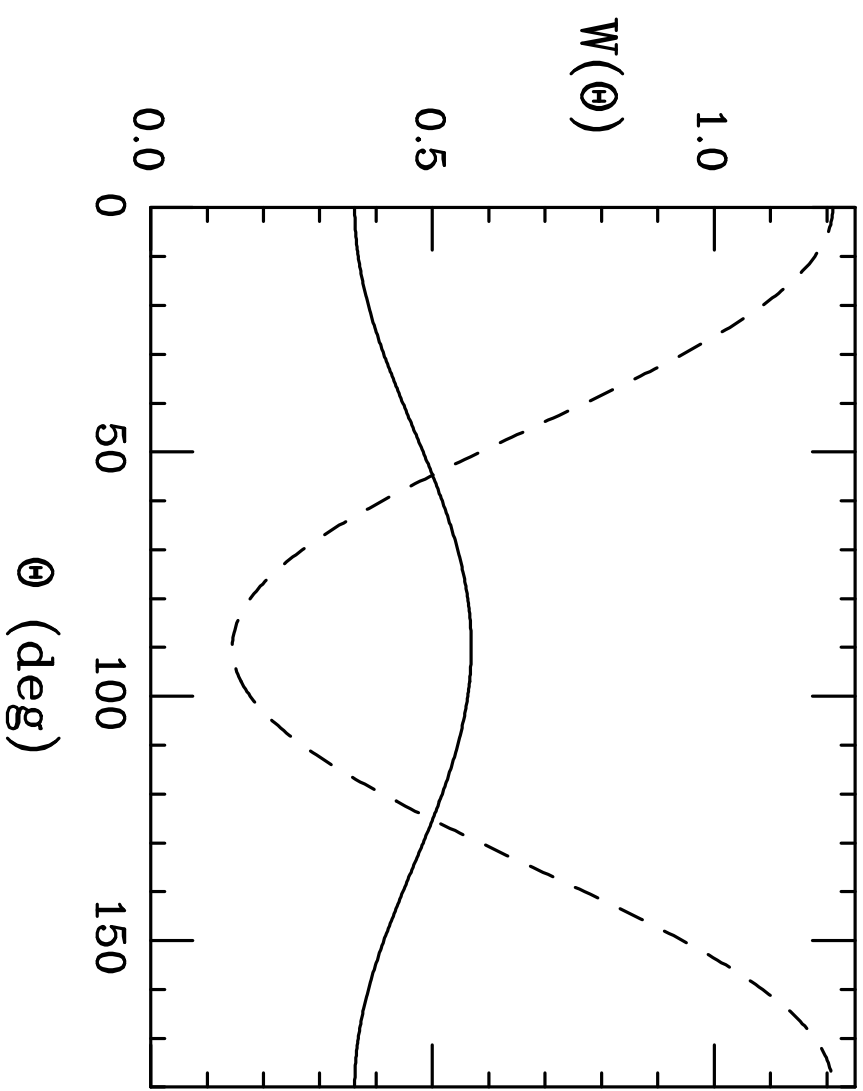
\begin{figure}
\caption[fig4]{ Plot of the angular distribution,
 $W(\Theta)$, of the $\pi$ emitted in the $\rho$ decay.
 The $\rho$ has been produced in the inclusive channel
 at an angle $\theta = 90^\circ$.
 Same notations for the curves as in Fig. 1. }
\label{win}
\end{figure}

\end{document}